\documentclass[letterpaper]{article} 
\usepackage{aaai2026}  
\usepackage{times}  
\usepackage{helvet}  
\usepackage{courier}  
\usepackage[hyphens]{url}  
\usepackage{graphicx} 
\usepackage{natbib}  
\usepackage{caption} 
\usepackage{algorithm}
\usepackage{algorithmic}

\usepackage{bm}
\usepackage{amssymb}
\usepackage{pifont} 
\usepackage{soul}
\usepackage{float}
\usepackage{array}
\usepackage{subfig}
\usepackage{amsmath}
\usepackage[table]{xcolor}
\usepackage[export]{adjustbox}
\usepackage{subfig}
\usepackage{subcaption}
\usepackage{verbatim}
\usepackage[nameinlink]{cleveref}
\usepackage{multirow}
\usepackage{multicol}
\usepackage{makecell}
\usepackage{minted}
\usepackage{adjustbox}
\usepackage{colortbl}
\usepackage{tabularx}
\usepackage{comment}
\usepackage{longtable}
\usepackage{fontawesome5}
\usepackage[acronym]{glossaries}
\glsdisablehyper
\newacronym{irb}{IRB}{Institutional Review Board}
\newacronym{ta}{TA}{Thematic Analysis}
\newacronym{ccm}{CCM}{Consolidated Cultural Model}
\newacronym{hci}{HCI}{Human-Computer Interaction}
\newacronym{ml}{ML}{machine learning}
\newacronym{ai}{AI}{Artificial Intelligence}
\newacronym{xai}{XAI}{Explainable AI}
\newacronym{emas}{EMAs}{Ecological Momentary Assessments}
\newacronym{nlp}{NLP}{Natural Language Processing}
\newacronym{cv}{CV}{Computer Vision}
\newacronym{lmm}{LMM}{Linear Mixed Model}
\newacronym{mse}{MSE}{Mean Square Error}
\newacronym{cscw}{CSCW}{Computer-Supported Cooperative Work \& Social Computing}
\newacronym{fnr}{FNR}{False Negative Rate}
\newacronym{fpr}{FPR}{False Positive Rate}
\newacronym{fdr}{FDR}{False Discovery Rate}
\newacronym{bh}{B-H}{Benjamini-Hochberg}
\newacronym{llms}{LLMs}{Large Language Models}
\newacronym{ubicomp}{Ubicomp}{Ubiquitous Computing}
\newacronym{psw}{PSW}{passive sensing for wellbeing}

\newcommand{\eg}{e.g.,\ }
\definecolor{green}{rgb}{0.08, 0.47, 0.16}
\definecolor{darkblue}{HTML}{0C0893} 
\definecolor{brilliantlavender}{rgb}{0.6, 0.4, 0.8}
\definecolor{candypink}{rgb}{0.89, 0.44, 0.48}
\definecolor{lightpurple}{rgb}{0.8, 0.5, 0.98}
\definecolor{green}{rgb}{0.08, 0.47, 0.16}
\definecolor{violet}{rgb}{0.96, 0.5, 0.5}
\definecolor{asparagus}{rgb}{0.53, 0.66, 0.42}
\definecolor{darkpastelpurple}{rgb}{0.59, 0.44, 0.84}
\definecolor{mediumslateblue}{rgb}{0.48, 0.41, 0.93}
\definecolor{darkpurple}{HTML}{4B569B}
\definecolor{lightred}{HTML}{E76547}
\definecolor{green1}{HTML}{719E65}
\definecolor{purple}{HTML}{4D5696}

\newcolumntype{L}[1]{>{\raggedright\let\newline\\\arraybackslash\hspace{0pt}}m{#1}}
\newcolumntype{C}[1]{>{\centering\let\newline\\\arraybackslash\hspace{0pt}}m{#1}}
\newcolumntype{R}[1]{>{\raggedleft\let\newline\\\arraybackslash\hspace{0pt}}m{#1}}
\newcolumntype{N}[1]{>{\raggedright\arraybackslash\setlength{\parindent}{0pt}}p{#1}}

\sethlcolor{cyan}

\newcommand\rec[1]{\textcolor{lightred}{M#1}}

\usepackage{newfloat}
\usepackage{listings}
\DeclareCaptionStyle{ruled}{labelfont=normalfont,labelsep=colon,strut=off} 
\floatstyle{ruled}
\newfloat{listing}{tb}{lst}{}
\floatname{listing}{Listing}
\title{Sources of Inequity and Fairness Risks in Wellbeing Sensing}
\author{
Han Zhang\textsuperscript{\rm 1}\thanks{This work was conducted during the author's PhD at the University of Washington.},
Vedant Das Swain\textsuperscript{\rm 2},
Koustuv Saha\textsuperscript{\rm 3},
Anind K. Dey\textsuperscript{\rm 4},
Jennifer Mankoff\textsuperscript{\rm 4}
}

\affiliations{
    \textsuperscript{\rm 1}University of Chicago\\
    \textsuperscript{\rm 2}New York University\\
    \textsuperscript{\rm 3} University of Illinois Urbana-Champaign\\
    \textsuperscript{\rm 4} University of Washington\\
    micohan@uchicago.edu, v.das.swain@nyu.edu, ksaha2@illinois.edu, anind@uw.edu, jmankoff@cs.washington.edu
}
\usepackage{bibentry}

\begin{document}

\maketitle

\begin{figure*}[t]
    \centering
\includegraphics[scale=0.36]{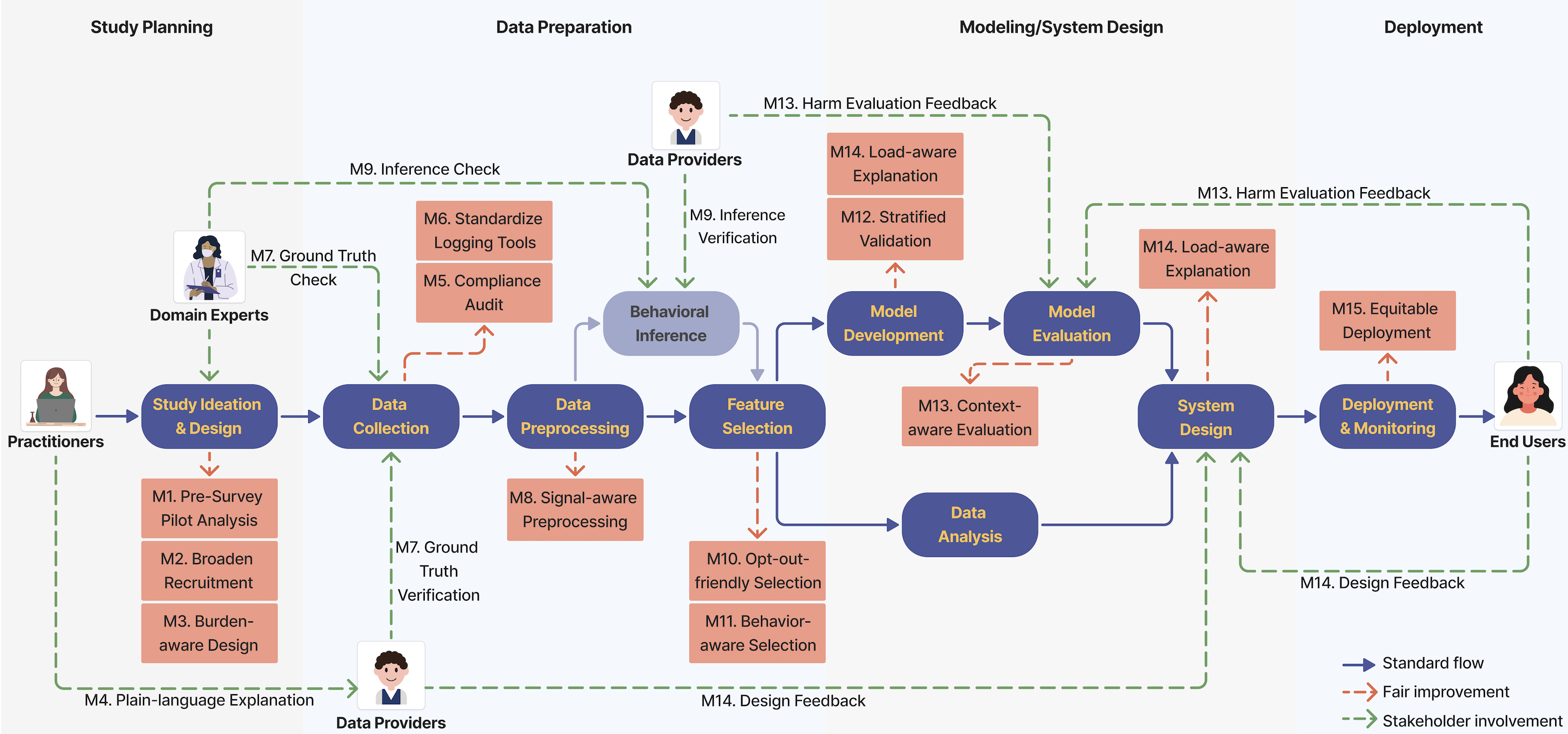}
    \caption{Overview of fairness risks and mitigation strategies across the passive sensing lifecycle. The lifecycle spans four phases from left to right: Study Planning, Data Preparation, Modeling/System Design, and Deployment. Arrows denote standard practices (\textcolor{darkpurple}{dark purple}), stakeholder involvement needed (\textcolor{green1}{green}), and proposed mitigation strategies (\textcolor{lightred}{red}). For readability, mitigation strategy labels (M1–M15) correspond to entries in Table~\ref{tab:risks_recommendations}. Behavioral inference is rarely an explicit pipeline stage in other ML systems.} 
    \label{fig:framework}
\end{figure*}

\begin{abstract}
Passive sensing for wellbeing uses smartphones and wearables to continuously collect human behavioral data and applies ML/AI models to infer psychological states and behaviors (e.g., depression, cognitive load). These systems are increasingly adopted in high-stakes settings (e.g., hospitals, universities), yet fairness research remains limited---primarily to post-hoc, identity-based comparisons of model performance. However, passive sensing combines heterogeneous sensing infrastructures, indirect behavioral inference, and longitudinal deployment---characteristics that, while not exclusive to the domain, are jointly pronounced here and raise two underexplored questions: (1) what additional sources of inequity arise from these characteristics, and (2) how do such inequities propagate beyond algorithmic audits across the system lifecycle? To address this gap, we conducted semi-structured interviews with 14 researchers and practitioners across five countries, examining how fairness risks emerge and are negotiated across the full passive sensing lifecycle. Our findings empirically characterize five situated sources of inequity (e.g., comfort with monitoring, behavioral regularity) that systematically shape fairness risks beyond identity-based attributes. We further synthesize 15 fairness risks and corresponding mitigation strategies across the lifecycle, from study design to deployment. Finally, we identify structural barriers that constrain fair practice in reality, and argue that enabling fair passive sensing requires both individual researcher efforts and ecosystem-level governance support from funders, publication venues, and deploying institutions.
\end{abstract}


\section{Introduction}

AI systems that continuously monitor human behavior and infer psychological states are no longer experimental. They are increasingly embedded in hospitals, workplaces, and universities as infrastructural tools for mental health monitoring~\cite{adler2020predicting,mohr2017personal}, workplace wellbeing~\cite{nepal2021assessing,arakawa2019sensing}, and educational support~\cite{wang2014studentlife,zhang2025towards}. Passive sensing systems---which collect behavioral data from smartphones and wearables to infer latent states such as stress, depression, and cognitive load~\cite{morshed2022advancing,jacobson2020passive,spathis2019passive,ahmadi2023cognitive}---exemplify this shift: from research prototype to deployed institutional technology operating at scale, across populations, and in high-stakes contexts.\footnote{Throughout this paper, we use \textit{passive sensing} to refer to applications in wellbeing-related domains, broadly construed to include mental and physical health and workplace wellbeing. Participants' examples occasionally draw on adjacent settings (e.g., productivity, care work); we retain these where they illustrate mechanisms relevant to wellbeing sensing.}

Yet as these systems move from design into deployment, their societal impact is shaped not only by technical choices but by who is included in data collection, who bears the burden of monitoring, and who has meaningful agency over how their behavioral data are interpreted and acted upon~\cite{selbst2019fairness,suresh2021framework}. The unequal distribution of these burdens and risks across individuals and populations is not merely a technical shortcoming---it is an ethical and governance failure that demands both researcher attention and institutional response~\cite{raji2020closing,madaio2020co}. Prior work has begun to examine fairness in passive sensing, most commonly through post-hoc comparisons of model performance across demographic groups (e.g., race, gender)~\cite{adler2024measuring,yfantidou2023uncovering,zhang2025towards,yfantidou2023beyond}. While valuable, these evaluations treat fairness primarily as a property of model outputs---obscuring how inequities arise earlier in the pipeline, accumulate across system stages~\cite{suresh2019framework}, and are shaped by institutional pressures that constrain what practitioners can do in practice~\cite{holstein2019improving,madaio2020co}.

To address this gap, we conducted semi-structured interviews with 14 researchers and practitioners across five countries---Australia, Canada, Japan, Korea, and the United States---examining how fairness risks emerge and are negotiated across the full passive sensing lifecycle. The cross-national scope of our study reveals how fairness risks are shaped not only by technical decisions but by the cultural, institutional, and organizational contexts in which passive sensing systems are built and deployed. Our findings reveal that inequities extend well beyond algorithmic bias: they are rooted in characteristics that are especially pronounced in passive sensing---heterogeneous sensing infrastructure~\cite{ferreira2015aware,mirjafari2019differentiating}, indirect behavioral inference~\cite{mohr2017personal,jacobs2021measurement}, and longitudinal deployment dynamics~\cite{rabbi2015mybehavior}---and it remains unclear how existing fairness frameworks should be adapted to address them. Importantly, researchers widely recognized these risks but described structural and institutional barriers---publication incentives, funding constraints, and organizational pressures---that systematically constrained their ability to act on fairness concerns in practice. This paper makes two contributions:

\begin{itemize}

   \item We empirically characterize five situated sources of inequity that shape fairness risks in passive sensing systems beyond identity-based attributes:  comfort with monitoring and vulnerability, device and sensor access, digital and data literacy, cultural or linguistic familiarity, and behavioral regularity.
    \item We synthesize 15 fairness risks and mitigation strategies across the passive sensing lifecycle (Table~\ref{tab:risks_recommendations}), and identify intervention points for different stakeholders (i.e., system builders, domain experts, data providers, and end users) (Figure~\ref{fig:framework}).
    
\end{itemize}

Together, these findings suggest that enabling fair passive sensing requires both individual researchers' efforts---such as conducting pilot analyses to surface situated inequities, designing studies with attention to burden on vulnerable populations (e.g., people with mental health concerns), and validating behavioral inference against participants' lived experiences (M1, M3, M9 in Figure~\ref{fig:framework})---as well as ecosystem-level governance support, including clearer expectations from funding bodies and publication venues, shared tooling and documentation standards, and institutional norms that reward transparency over performance.


\section{Background and Related Work}

\subsection{Passive Sensing in Wellbeing}

Passive sensing uses smartphones and wearables (e.g., Apple Watch, Oura Ring) to continuously and automatically collect behavioral data from individuals in everyday life---without requiring deliberate user input~\cite{harari2016using,onnela2021opportunities}. Common data streams include heart rate, screen time, sleep patterns, physical activity, and location traces~\cite{wang2014studentlife,zhang2025towards,adler2020predicting}. These behavioral signals are then used to infer human behaviors and psychological states---such as stress, depression, and cognitive load---using ML/AI models~\cite{mohr2017personal,jacobson2020passive}.

A passive sensing lifecycle may proceed through up to four stages: (1)~\textit{study planning}, where researchers define target populations and select sensing modalities; (2)~\textit{data preparation}, where raw sensor streams are cleaned, transformed into interpretable behavioral constructs, and prepared for modeling; (3)~\textit{modeling}, where \gls{ml}/AI models are trained to predict outcomes from those constructs; and (4)~\textit{deployment}, where outputs are surfaced to end users or decision-makers~(Figure~\ref{fig:framework}). 

Stages 2 and 3 are conceptually distinct in ways that matter for fairness. Behavioral inference, characteristic of passive sensing pipelines, involves transforming raw sensor signals into meaningful behavioral constructs through rule-based or signal-processing methods---for example, converting raw GPS coordinates into \textit{location entropy} as a proxy for social isolation~\cite{canzian2015trajectories}. These constructs are not predictions; they are intermediate representations of latent behavioral patterns. Modeling, by contrast, involves training statistical or \gls{ml} models to \textit{predict} a clinical or behavioral outcome---such as depression severity or stress level---from the behavioral constructs derived in stage 2~\cite{jacobson2020passive}. This distinction matters for fairness: inference errors in stage 2 (e.g., a behavioral construct that carries different meaning across cultural contexts) propagate silently into stage 3, where they manifest as model errors that are difficult to diagnose without examining the upstream 
pipeline~\cite{suresh2021framework}.

\subsection{Fairness-Relevant Characteristics of Passive Sensing}
Passive sensing concentrates three characteristics that shape how fairness risks arise. None is exclusive to the domain: educational assessment, for example, also combines heterogeneous measurement (teachers' records of varying quality), inferred latent constructs (a silent student may have mastered the material or be lost), and longitudinal data collection. Our claim is narrower---these characteristics are jointly pronounced in wellbeing sensing, and their fairness implications there remain empirically underexamined.

First, passive sensing  operates within \textbf{heterogeneous sensing infrastructure}---including device type, operating system, and environmental context---that systematically affects what data are captured and how they should be interpreted~\cite{shcherbina2017accuracy,mirjafari2019differentiating}. For example, differences in sensing capabilities between iOS and Android devices are well documented~\cite{harari2016using,onnela2021opportunities}, and smartphone ownership patterns correlate with socioeconomic status, with iOS users on average having higher income than Android users~\cite{jamalova2019comparative}. Design choices that rely on device-specific data may therefore disproportionately affect lower-income populations, even when model performance appears similar across demographic groups.

Second, passive sensing relies on \textbf{indirect behavioral inference}: unlike most AI/\gls{ml} tasks with directly observed labels, passive sensing often infers internal states (e.g., stress, mood, cognition load) from behavioral proxies (e.g., screen time or mobility regularity) whose meaning is not intrinsic to the signal but emerges from individuals' situational, social, and cultural contexts~\cite{mohr2017personal,onnela2021opportunities,cornet2018systematic}. For example, the same behavioral signal (e.g., phone calls) may indicate anxiety for one person but social connection for another~\cite{assi2023complex,meegahapola2023generalization}. As a result, errors may arise not only from model miscalibration but from mismatches between assumed and actual behavioral meaning that are difficult to detect through standard evaluation.

Third, passive sensing's  \textbf{longitudinal and intervention-oriented} data collection and deployment create cumulative and unevenly distributed burden. Populations already facing higher monitoring demands---such as people with mental health conditions~\cite{ben2016mhealth,huckins2020mental} and students under academic stress~\cite{wang2015smartgpa,zhang2025towards}---may be disproportionately affected when sensing systems increase surveillance or skew representation~\cite{ben2016mhealth,canali2024wearable}. In deployment, predictions often trigger behavioral nudges that create feedback loops, complicating fairness evaluation and long-term reliability~\cite{mohr2017personal,rabbi2015mybehavior,adanyin2024ai,matthews2014tracking}. 

Existing fairness tooling already covers parts of this space: standard group audits, for instance, can treat device type or operating system as the comparison attribute and test for performance disparities. What such tools leave open---and what motivates this study---is how they should be adapted and evaluated for risks such as unverified behavioral inference, cumulative monitoring burden, and consent under seamless longitudinal collection.


\subsection{Definition of Fairness and Fairness Audits in Passive Sensing}

In \gls{ml}, fairness has most commonly been operationalized as comparable performance across individuals or groups, evaluated through metrics such as differences in error rates, AUC, or demographic parity across predefined attributes like gender or race~\cite{chouldechova2017fair,pleiss2017fairness,mehrabi2021survey}. Recognizing the socio-technical nature of algorithmic systems, a growing body of research has argued that fairness should instead be understood as an emergent property of systems that intertwine technical components with social practices, institutional structures, and human values~\cite{selbst2019fairness,suresh2021framework,barocas2016big,katell2020toward}. Our definition---the absence of systematic inequities in who is represented and has access, who bears burden and error, and who has meaningful agency---extends prior work by treating representation, burden, and procedural agency as fairness-relevant properties across the full sensing lifecycle, consistent with sociotechnical critiques of measurement~\cite{jacobs2021measurement}, lifecycle harms~\cite{suresh2019framework,suresh2021framework}, and algorithmic abstraction~\cite{scheuerman2019computers}.

As passive sensing has moved into real-world use, a growing body of work examines fairness through model audits comparing predictive performance across identity-based attributes such as age, gender, race, and health condition~\cite{yfantidou2023uncovering,adler2024measuring,zhang2023framework,chancellor2019taxonomy,mader2024learning,khwaja2019modeling,assi2023complex,chowdhary2023can}. These studies document meaningful disparities in high-stakes domains---for example, higher risk scores assigned to certain racial minority participants~\cite{adler2024measuring} or disproportionate misclassification of first-generation students ~\cite{zhang2025towards}---and are essential for surfacing inequities that would otherwise remain hidden. At the same time, recent work highlights the limits of identity-based evaluations: models trained in one setting often degrade when applied elsewhere, and qualitative work has raised concerns about surveillance creep, misaligned wellbeing definitions, and uneven consent and contestability in sensing deployments~\cite{kawakami2023sensing,chowdhary2023can,meegahapola2023generalization}. 

Despite this progress, fairness research in passive sensing remains largely centered on post-hoc model audits, leaving less attention to how fairness risks arise upstream and accumulate across the system lifecycle---and still less to how domain practitioners reason about and attempt to mitigate these risks under real institutional constraints.

\subsection{Lifecycle-awareness of Fairness in ML and HCI} 

Building on a socio-technical view of fairness, prior work in \gls{ml} and \gls{hci} demonstrates that many fairness-related harms originate upstream of model development~\cite{suresh2021framework,paullada2021data,sambasivan2021everyone,jacobs2021measurement,liu2023reimagining,kim2025systematic,holstein2019improving}. Early decisions about problem framing, objectives, and feature selection can pre-commit systems to particular notions of ``success'' that propagate through later stages of development and deployment~\cite{veale2018fairness,passi2019problem,binns2018fairness,birhane2022values}. In practice, limited visibility into data provenance and representativeness further constrains practitioners' ability to anticipate which populations a system may underserve~\cite{holstein2019improving,raji2020closing,gebru2021datasheets,mitchell2019model}. Dataset composition and labeling practices can also encode structural bias, illustrating how recruitment and annotation choices shape downstream error patterns~\cite{scheuerman2019computers,buolamwini2018gender,hanna2020towards}. Once embedded in data and workflows, these choices are difficult to remediate through post-hoc thresholds or model-only fairness metrics~\cite{suresh2021framework,paullada2021data,madaio2020co,green2020algorithmic}.

Downstream, deployment and use further influence how fairness manifests in practice. Empirical audits show that disparities often surface during real-world use, and that addressing them requires ongoing organizational processes and oversight rather than one-time metric checks~\cite{raji2019actionable,raji2020closing}. In response, fairness governance research has proposed a range of lifecycle-oriented tools, including Datasheets for Datasets and Model Cards to document data provenance and intended use~\cite{gebru2021datasheets,mitchell2019model,hutchinson2021towards}, co-designed checklists that support role-specific decision-making~\cite{madaio2020co,holstein2019improving,veale2018fairness}, Human–AI interaction guidelines for explanation and recourse~\cite{amershi2019guidelines,liao2020questioning}, and internal or third-party audits that link evaluation to organizational accountability~\cite{raji2019actionable,raji2020closing}. While these tools provide important foundations for fairness governance, they have largely been developed in domain-general settings. Without empirical grounding in the practices and constraints of passive sensing, it remains unclear when, how, and to what extent such lifecycle-oriented approaches can be effectively adapted, operationalized, or evaluated in sensing-based systems.

\section{Interview Study}\label{sec:interviews}

We conducted semi-structured interviews with researchers and practitioners working on passive sensing systems to (1) understand how domain-specific characteristics shape fairness in practice, and (2) examine where fairness risks arise across the sensing lifecycle, as well as the mitigation strategies and barriers that shape how these risks are addressed.

\subsection{Recruitment \& Participants} We recruited participants using purposive and network-based (snowball) sampling over a one-month period. We first contacted individuals in our professional networks with experience conducting sensing studies in academic or industry settings, then expanded the pool by inviting referrals to broaden perspectives across countries and levels of expertise. Inclusion criteria required participants to be at least 18 years old and to have designed or conducted at least one passive sensing study that developed predictive models from sensing data to infer human behaviors or internal states (e.g., wellbeing, workplace productivity, and social interaction). Each participant received an Amazon gift card as compensation: \$40 for sessions lasting 60–85 minutes and \$60 for sessions exceeding 85 minutes. The study was approved by our \gls{irb}.

Interested participants completed a brief pre-survey, administered via Google Form, collecting  demographic information such as gender, current role, years of experience, and primary passive sensing application domains. The survey also asked for participants' initial perspectives on fairness-related concerns and what challenges limited such consideration (see in Appendix). These responses helped inform and tailor our semi-structured interview protocol. In total, we interviewed 14 researchers and practitioners from five countries (Australia, Canada, Japan, Korea, and the U.S.). All had experience conducting sensing-based studies related to health and wellbeing, and several also had experience in adjacent domains (e.g., social behavior, human interaction or workplace productivity). Table~\ref{tab:participant_summary} in the Appendix summarizes key participant demographics and study identifiers. The pool skews toward academia: eleven participants held academic roles (seven PhD students, two postdoctoral researchers, two faculty), and three were industry researchers (Table~\ref{tab:participant_summary}). Our findings therefore primarily characterize research practice; we return to this in Limitations. 

\subsection{Procedure} We conducted each interview remotely over Zoom. One author led each session and, with participant consent, recorded audio/video solely for verbatim transcription. Participants could decline any question, ask for clarification, and remove sensitive content at the end of the session. The full interview instrument appears in the Appendix.

Interviews proceeded in three parts. In Part 1, participants described a prior passive sensing study and mapped its workflow on a shared FigJam whiteboard~\cite{figma}, then articulated how they define fairness in passive sensing and whether it extended beyond model performance. We probed experiences with behavioral inference, including mismatches between inferred behaviors and participants' actual situations, and asked about barriers to fairness consideration. In Part 2, participants identified lifecycle stages introducing fairness risks, described methods and metrics used to assess and mitigate them, and reflected on how longitudinal data dynamics shaped their approaches. In Part 3, participants revisited their pipeline diagrams to annotate fairness risks and mitigation strategies stage by stage, then provided feedback on a draft lifecycle-aware fairness framework adapted from prior work~\cite{zhang2023framework}.

\subsection{Data Analysis}
We transcribed all interviews, wrote an analytic memo after each session, and documented---after each interview---any artifacts that participants had shared during or after their sessions. To ensure participant confidentiality, we removed names and other identifying information from all transcripts. Interviews ranged from 70 to 110 minutes (mean = 86, SD = 13); two participants completed their interviews in two sessions due to scheduling constraints. We analyzed the transcripts using \gls{ta}~\cite{braun_using_2006} , following the five-phase process outlined by \citeauthor{braun_using_2006}. One author began by reading all transcripts to gain familiarity with the data and drafted an initial codebook grounded in the interview structure and content. Three rounds of collaborative coding followed, during which two authors independently coded the same transcript and resolved discrepancies through a mix of synchronous (Zoom) discussions and asynchronous dialogue. After establishing consistency, one author coded the remaining transcripts in two iterative passes, allowing for refinement and expansion of the codebook as new patterns emerged. The other author randomly selected and reviewed three of the coded transcripts to provide feedback and ensure continued alignment. Throughout, we maintained detailed documentation of analytic decisions and codebook iterations to support rigor and transparency. 

\section{Findings}

We organize our findings around three themes comprising distinct yet interrelated subthemes. Many issues span multiple themes and lifecycle stages. We use \textit{researchers} to refer to our interviewees and \textit{participants} to refer to data providers and/or end users in the studies they described.

\subsection{Situated Sources of Inequity in Passive Sensing}\label{subsubsec:fairness_in_wellbeing_sensing}

Across interviews, researchers identified five recurring, situated sources of inequity in passive sensing that extend beyond identity-based attributes: (1) comfort with monitoring and vulnerability, (2) device and sensor access, (3) digital and  data literacy, (4) cultural or linguistic familiarity, and (5) behavioral regularity. These sources do not operate in isolation. Instead, they shape fairness by differentially affecting who is represented in data, who bears monitoring burden and inference errors, and who has meaningful procedural agency throughout passive sensing studies. We organize these findings into three sub-themes that capture both the origins and downstream consequences of unfairness.

\subsubsection{Representation \& Access.}
Beyond identity-based attributes, several researchers raised concerns about participants' comfort with continuous, passive monitoring in longitudinal studies, noting that it shaped both enrollment and retention. For privacy-protective, highly conscientious participants with low institutional trust, always-on sensing was especially off-putting, often reducing participation over time. This pattern introduces self-selection bias that disproportionately excludes vulnerable populations---often those most in need of support---resulting in models trained on unrepresentative data and likely to fail for high-burden groups. As P8 observed,  the individuals their studies aim to understand and support are often ``\textit{conscientious people with low trust in institution.}'' Researchers further described concrete triggers that led participants to withhold permissions or drop out mid-study---\eg ``\textit{continuous GPS traces}'' and ``\textit{video capture}'' in private settings. For example, in a study of workers' mental wellbeing, one researcher noted: ``\textit{I can clearly explain what workers didn't like. They didn't want to their GPS [data] to be collected}'' (P14).

In addition to participation preferences, researchers re-emphasized concerns about inequities in device and sensor access. Platform-locked studies (\eg iOS-only deployments) systematically exclude Android users, a constraint that often intersects with socioeconomic status. As P11 noted, such exclusions can effectively filter out participants who are ``\textit{probably socioeconomically not that [well off],}'' resulting in systematic data gaps. Beyond smartphones, reliance on specialized or uncommon sensors (\eg Oura Ring) further narrowed participation; as P7 cautioned, such design choices bias samples toward the device's typical user base. Researchers also pointed to disparities in digital literacy as a persistent source of exclusion: participants who were ``\textit{less technologically prone or have less digital literacy}'' (P11) were more likely to be underrepresented, disengage over time, or contribute incomplete data. 

Cultural and linguistic familiarity also introduced fairness issues in representation and engagement. Researchers described how study materials grounded in local sociopolitical contexts could unintentionally exclude participants without that background knowledge. For example, P12 discussed a study conducted in Australia that used locally specific stimuli (\eg debates on abortion and same-sex marriage),  which many international students ``\textit{didn't get}'' due to lacking familiarity with Australian societal contexts. Even proficient English speakers from international backgrounds were ``\textit{not familiar with the debates},'' resulting in disengagement and uneven task participation (P12). This systematically excluded culturally unfamiliar participants from meaningful inclusion, and undermined the validity of inferences drawn from disengaged or confused respondents.

\subsubsection{Burden \& Error.} Once included in a study, participants encountered unevenly distributed burdens and risks that extended beyond data collection to encompass ground-truth generation, inference accuracy, and the downstream consequences of model error. Researchers consistently emphasized how differences in digital literacy shaped the disproportionate burden of producing ground truth, particularly in studies relying on high-frequency self-report measures. Persistent notifications and text-heavy prompts often fatigued participants, increasing disengagement over time. As P11 observed, text-heavy, long-form prompts were primarily tolerated by technically adept participants (\eg ``\textit{computer science students}''), whereas others often disengaged due to cognitive overload---resulting in noisier or incomplete ground truth from non-technical participants and biasing models toward privileged behavioral patterns. 

Beyond ground truth quality, researchers emphasized that cultural context can introduce systematic inference and model errors. They described cases such as consistently high productivity despite poor mental-health reports among Japanese workers (P9), or gentle typing behaviors under stress among Korean office workers that complicated attempts to infer anger (P14). Behavioral regularity further shaped inference and model accuracy: participants with highly regular routines often yield ``\textit{easier}'' insights and more stable predictions, whereas those with irregular or constrained routines were more likely to be mischaracterized. 

Moreover, vulnerability introduced compounded burdens, including physical safety risks when sensors were misidentified by older participants---for example, mistaking a bed-side beacon for food (P9). These burdens further interacted with how system outputs or interventions were delivered. Because human internal states are ``\textit{so private and intimate}'' (P2), delivering an incorrect prediction (\eg a false positive)---or delivering it poorly (ill-timed, little explanation)---to someone already in a distressed or vulnerable state can amplify anxiety, erode trust, or trigger harmful rumination. As P6 warned: ``\textit{If a depressed person sees, `Oh, your expression is very depressed,' that result can exacerbate the situation. We can see a worsen problem here.}''

\subsubsection{Procedure \& Recourse.} Procedural unfairness emerged when participants lacked meaningful ability to understand, contest, or control how their behavioral data were collected and interpreted. Researchers consistently described this as an issue of asymmetric data and digital literacy, rather than a failure of formal consent mechanisms alone. 

Several researchers noted that passive sensing relies on inferences that are difficult for participants to anticipate, making consent challenging even when opt-out options are provided. As P2 explained, behavioral signals ``\textit{are not a weight, nor a size,}'' meaning their implications are neither transparent nor intuitively understood. Participants may therefore agree to data collection without grasping the scope of what can be inferred. They further reflected, although participants are told they can decline sharing data, ``\textit{if someone underestimates how powerful data is, they will not use this right,}'' resulting in what researchers characterized as procedural agreement without substantive understanding. This gap led to situations where researchers ``\textit{get a lot of data even though the people might not want to share it, actually, if they knew what we learned from them.}'' This challenge was compounded by the seamless and longitudinal nature of passive sensing. Researchers questioned whether participants could remain aware of ongoing data collection and inference over time, particularly when sensing occurred in the background. As P12 asked, ``\textit{Do we inform the users that their signals are being collected, given that passive sensing can be very seamless and progressive?}'' 

Downstream, procedural unfairness also manifested through unequal digital or data literacy. Researchers observed that participants with lower literacy often lacked clear explanations for why they received particular messages or interventions, leading to confusion or disengagement (P11).  While researchers emphasized the normative importance of transparency---``\textit{to do everything to give all participants the information they need to make a decision}'' (P3)---they acknowledged that, in practice, only some participants were able to act on this information. Those with higher literacy or professional backgrounds were more likely to question, interpret, or challenge system outputs, whereas others were more likely to accept decisions without understanding or withdraw from participation altogether.


\subsection{Fairness Risks Across the Passive Sensing Lifecycle}\label{subsec:fairness_risks}

\begin{table*}[!t]
\sffamily
\caption[Identified fairness risks and mitigation strategies across the passive sensing lifecycle.]{Identified fairness risks and mitigation strategies across the passive sensing lifecycle.}
\label{tab:risks_recommendations}
\centering
\scriptsize
\setlength{\tabcolsep}{0.9mm}{
\begin{tabular}{|m{0.035\textwidth}|m{0.135\textwidth}|p{0.32\textwidth}|p{0.025\textwidth}|p{0.4\textwidth}|}
\hline 
\multicolumn{1}{|c|}{\textbf{Phase}} & \multicolumn{1}{c|}{\textbf{Stage}} & \multicolumn{1}{c|}{\textbf{Potential Risks}} & \multicolumn{1}{c|}{\textbf{\#}} & \multicolumn{1}{c|}{\textbf{Mitigation Strategies}} \\\hline

\parbox[t]{2mm}{\multirow{7}{*}{\rotatebox[origin=c]{90}{Study Planning}}} & 
\multirow{7}{*}{Study Ideation/Design} & Over-reliance on identity-based audits and limited attention to situated inequities can mis-specify comparison groups. & \rec1 & Pilot analyses to identify situated factors (e.g., monitoring comfort, digital literacy) predicting dropout or data gaps; use as stratification criteria alongside identity attributes.
 \\\cline{3-5}
& & Convenience sampling excludes target subgroups, introducing representativeness gaps. & \rec2 & Set explicit inclusion targets for relevant situational subgroups beyond convenience samples. \\\cline{3-5}
& & Poor study design can disproportionately burden certain participants and yield uneven data. & \rec3 & Calibrate burden via pilots; tune survey frequency, length, and question order. \\\hline

\parbox[t]{2mm}{\multirow{19}{*}{\rotatebox[origin=c]{90}{{Data Preparation}}}} & \multirow{8}{*}{Data Collection}
 & Uneven digital/data literacy at consent undermines informed decision-making about data sharing. & \rec4 & Plain-language consent with examples, comprehension checks, and standardized ethical checklists. \\\cline{3-5}

&  & Comfort with monitoring affects compliance in longitudinal collection, creating systematic data gaps. & \rec5 & Iteratively audit compliance to surface disparities; minimize burden to sustain engagement. \\\cline{3-5}
&  & Data logging tool heterogeneity introduces systematic differences across participant data. & \rec6 & Standardize or document software tools, versions, and capabilities to ensure comparability. \\\cline{3-5}
& & Proxy annotators and rushed self-reports create uneven ground truth quality. 
& \rec7 
& Validate labels with participants; involve domain experts for adjudication; use multiple annotators. \\\cline{2-5}
& \multirow{3}{*}{Data Preprocessing} & Context-insensitive preprocessing (e.g., normalization, imputation) erases meaningful subgroup variation. & \rec8 & Apply preprocessing with attention to subgroup variation; avoid compressing meaningful behavioral differences into indistinguishable values. \\\cline{2-5}
& \multirow{3}{*}{Behavioral Inference} & Unverified behavioral inference misrepresents participants' lived experiences. & \rec9 & Validate inference by presenting sampled episodes to data providers or domain experts; avoid inference when validity cannot be established. \\\cline{2-5}
& \multirow{5}{*}{Feature Selection} & Uneven comfort with sensitive data types leads to coverage gaps and performance disparities. & \rec{10} & Explain data use and necessity; design models to flexibly exclude sensitive features without substantial performance loss. \\\cline{3-5}
&  & Feature selection optimized for aggregate accuracy excludes contextually meaningful signals, disadvantaging atypical participants. & \rec{11} & Examine underperforming features before discarding; use explainability tools to assess whether exclusion disproportionately obscures meaningful signals. \\\hline

\parbox[t]{2mm}{\multirow{6}{*}{\rotatebox{90}{\raisebox{1.5em}{{\makecell[c]{Modeling \\ /System Design}}}}}} & \multirow{2}{*}{\makecell[l]{Model Development}} & Models trained on small, uneven datasets generalize poorly across subgroups. & \rec{12} & Stress-test via multiple validation schemes; use subgroup-specific ensembles; choose model families matching data scale. \\\cline{2-5}
& \multirow{2}{*}{Model Evaluation} & Misaligned metrics and absent participant-centered verification mask unfair system behavior. & \rec{13} & Align metrics with task characteristics; incorporate participant judgments; use visualizations to monitor uneven performance. \\\cline{2-5}
& \multirow{2}{*}{\makecell[l]{Model Development  \\ \& System Design}} & Explanations are received unevenly across groups due to cultural and literacy differences. & \rec{14} & Account for cultural context and information intensity; incorporate participant feedback into explanation design. \\\hline
\parbox[t]{2mm}{\multirow{2}{*}{\rotatebox[origin=c]{90}{{Depl.}}}} & \multirow{2}{*}{Deployment} & Representativeness, access, or affordability barriers reproduce upstream bias. & \rec{15} & Promote equitable access; monitor post-deployment performance across groups to surface and address emerging disparities. \\\hline
\end{tabular}}
\end{table*}

Across interviews, researchers emphasized that fairness risks in passive sensing rarely arise at a single point; instead, they accumulate and transform across the system lifecycle. We therefore synthesize these insights into a set of recurring fairness risks paired with corresponding mitigation strategies proposed by researchers (Table~\ref{tab:risks_recommendations}), and highlight potential actions by different stakeholders across the lifecycle (Figure~\ref{fig:framework}). 

\noindent \textbf{Study Planning}.
At the study ideation stage, researchers identified one early fairness risk: limited attention to situated sources of inequities and an over-reliance on identity-based audits that mis-specify comparison groups. These gaps can introduce fairness risks before data collection begins and propagate through subsequent stages. For example, researchers cautioned that audits solely on identity-based attributes may fail to capture behavioral variation, leading to downstream choices grounded in flawed assumptions.  As P7 noted, ``\textit{demographic subgroups}'' do not reliably determine how behavior relates to mental health; instead, relevant subgroups may be defined by shared behavioral–outcome relationships that are highly individual and dynamic. To mitigate these risks, researchers emphasized conducting pilot analyses  to assess whether monitoring comfort, behavioral regularity, or digital literacy predict dropout or data quality gaps, then using these factors as stratification criteria alongside identity attributes (P1–P5, P7–P11, P14; \rec{1}).

In practice, many studies relied on university populations because they are easier to approach and recruit. However, this sampling often produced fundamental representation gaps, as such samples were poor proxies for intended participants. For example, one study using an international-predominant university sample failed to reflect the local Western population under investigation (P12), while research targeting older adults with dementia showed that university-centric recruitment can miss the target population entirely (P4, P9). This early exclusion constrains representativeness from the outset and limits the real-world validity of resulting models. As P4 observed, ``\textit{it is absolutely not possible to detect the movement of dementia patients if the model is trained on younger adults}.'' To address this, researchers recommended broadening recruitment beyond convenience samples by setting explicit inclusion targets for relevant situational subgroups (P2, P5, P9–P12; \rec{2}).


Researchers further cautioned that study design choices (\eg sensor selection and self-report construction) can disproportionately burden certain participants and yield data that unevenly reflect participants' lived realities (\eg older patients with dementia). P3 noted that high-frequency, lengthy \gls{emas}---brief, repeated surveys sent to participants throughout the day---encourage ``\textit{satisficing,}'' where participants complete surveys with patterned or low-effort responses, degrading data quality. To calibrate burden, researchers recommended pilots to estimate the minimum data required and design controls such as randomizing question order and carefully tuning survey frequency and length to reduce response artifacts (P3, P9; \rec{3}).


\noindent \textbf{Data Preparation}. Once inclusion criteria were set, researchers identified uneven digital and data literacy at the point of consent as a core fairness risk. Participants often lacked sufficient understanding to make informed decisions about data sharing, undermining meaningful consent. To address this, researchers emphasized plain-language explanations of what data are collected and why---often supported by certain examples---along with clearly stated study goals, and brief comprehension checks such as ``\textit{short questionnaires}'' (P4). They further called for transparency around specific data uses, process-level supports such as ``\textit{standardized fairness and ethical checklists}'', and more accessible and interpretable consent materials for {data providers} with different literacy levels (P2, P9, P10, P12; \rec{4}).

When discussing fairness risks in data collection, researchers reiterated that limited attention to domain-specific context and narrow sampling can introduce representativeness bias (P2, P4–P5, P7–P12). For example, they noted that compliance in longitudinal studies declined unevenly based on participants' comfort with monitoring and individual traits (P5, P8, P10–P11), creating systematic gaps in data coverage over time. To mitigate this risk, researchers recommended iteratively auditing compliance to surface emerging disparities and minimizing participant burden to sustain engagement  (P1–P5, P7–P10, P12, P14; \rec{5}). In addition, variation in logging tools and third-party software introduced systematic differences in data quality (P3). Researchers therefore emphasized standardizing or carefully documenting software tools, versions, and capabilities to ensure data comparability (P7, P8; \rec{6}).

When participant-reported ground truth was infeasible, some studies relied on proxy annotators (e.g., caregivers). Researchers cautioned that real-world constraints make such labels systematically noisier for certain groups—for instance, caregivers prioritizing physical care introduced temporal misalignment in activity logs (P9), and hurried self-reports compound these effects (P3). These gaps are not evenly distributed, disproportionately degrading ground-truth quality for groups requiring proxy labeling. To mitigate these risks, they recommended validating labels with {data providers} when feasible, incorporating {domain experts} for adjudication, and using multiple annotators to cross-check or triangulate labeled events (P3–P4, P6, P9; \rec{7}).

While these strategies can improve data collection and labeling, researchers emphasized that fairness risks persist during data preprocessing. Common techniques such as normalization or imputation, when applied without contextual consideration, can distort or erase meaningful subgroup variation and individual differences. As P10 noted, normalization may compress distinct behavioral patterns into indistinguishable values, causing signals that differ in practice to ``\textit{all mean the same thing}'' depending on how preprocessing is performed (\rec{8}). Such choices risk biasing downstream interpretation and obscuring the heterogeneity that passive sensing systems seek to capture.

In contrast to other stages---where fairness concerns were often raised unprompted---risks associated with behavioral inference surfaced primarily when explicitly probed. Researchers frequently treated inference as a routine extension of preprocessing or feature engineering, and did not initially articulate it as a distinct fairness risk. However, when asked directly about mismatches between inferred behaviors and participants' lived experiences, all researchers who performed inference acknowledged such discrepancies. Despite recognizing these mismatches, only two researchers reported verifying or correcting inferred behaviors prior to modeling, typically by cross-checking with self-reports. Others cited barriers including limited ``\textit{domain knowledge}'' to interpret discrepancies (P1), the insufficiency of ``\textit{quantitative}'' signals alone to explain behavior (P8, P11), uncertainty about whether participants' reports reflected the ``\textit{actual situation}'' (P11, P12), and the difficulty of intervening in ``\textit{real-world settings}'' where inference occurs (P14). To mitigate these risks, researchers proposed complementing quantitative inference with {data providers'} brief qualitative debriefs and {domain expert} consultation---presenting a random sample of inferred behavioral episodes to domain experts or to data providers and ask whether the inferences match their recollection---or, when validity could not be reasonably established, avoiding inference altogether in favor of raw sensing features (P3, P5–P7, P9; \rec{9}).

At the feature selection stage, the choice of sensing modalities introduced uneven fairness risks across groups. Researchers consistently flagged video- and camera-based data as especially sensitive, noting that discomfort with visual recording was not evenly distributed (P1, P3, P8–P10, P12–P14). Many of them noted that during their studies, even when video was collected, some participants expressed unease about potential future viewing or secondary analysis. To mitigate these risks, researchers emphasized transparency about how each data type would be used and why it was necessary (P3, P6; \rec{10}). Several also advocated for designing models that can flexibly exclude certain features---either at participants' request or for fairness reasons---without substantially degrading performance (P3, P6; \rec{10}). However, one researcher acknowledged that building such flexibility into systems remains a technical challenge (P3).

Beyond sensitive data use, researchers identified feature selection as a key source of fairness risk. Optimizing features primarily for aggregate accuracy or interpretability can exclude contextually meaningful signals (P4–P6, P9–P11). For example, P4 noted that features that underperform at the population level may still encode important behavioral variation; removing them can yield models that perform well on average while systematically failing participants whose behaviors deviate from dominant norms. Rather than discarding such features completely, researchers suggested carefully examining the underlying reasons for their performance, using explainability tools, and considering whether their exclusion could disproportionately obscure meaningful behavioral signals (P4, P9, P14; \rec{11}).

\noindent \textbf{Modeling and System Design}. Researchers raised concerns about limited generalizability during model development: models often performed well for some subgroups but poorly for others (P1, P3–P5, P7, P10, P12). Beyond representativeness issues inherited from data collection, the small sample sizes typical of passive sensing further constrain generalization across contexts and populations---limitations that, as P3 noted, are often underreported. To mitigate these risks, researchers recommended stress-testing models through multiple validation schemes, training subgroup-specific models that can be combined via ensembles, and selecting model families appropriate to data scale to avoid overfitting---\eg avoiding overly ``\textit{complex deep learning techniques}'' for small samples (P4, P5, P7; \rec{12}). 

When evaluating models, researchers highlighted two recurring concerns: misaligned metric choice and the absence of participant-centered verification (P1, P2, P5–P7, P11). Default metrics such as accuracy or F1 were often applied without considering whether they capture key characteristics of passive sensing data (\eg situational). Researchers also cautioned that fairness assessments relying solely on quantitative scores can overlook whether system outputs are experienced as ``\textit{fair or appropriate}'' by participants themselves. To mitigate these risks, participants advocated aligning evaluation metrics with the properties of behavioral data---explicitly accounting for context and class imbalance---and using visualizations to monitor anomalous or uneven performance (P1, P6, P11; \rec{13}). Several also called for ``\textit{clearer evaluation guidelines}'' on metric selection and for incorporating qualitative feedback from participants---including {data providers} and {end users}---alongside quantitative evaluation (P1, P2, P6, P7; \rec{13}).

For model development and system design, researchers emphasized that gaps in explanation are unevenly experienced across participants. They critiqued the slow uptake of explainability practices in passive sensing, noting that systems are often designed as black boxes and, when interpretable, are primarily explainable to researchers rather than to participants themselves (P2, P4, P9, P11, P14). Moreover, researchers noted that explanation practices vary widely in effectiveness, shaped by ``\textit{cultural differences}'' in how explanations are perceived, the need to calibrate ``\textit{information intensity}'' to avoid overwhelming participants, trade-offs between detail and engagement, and whether participant feedback---from both {data providers} and {end users}---is incorporated into explanation design (P3, P4, P9-P11; \rec{14}). 


\noindent \textbf{Deployment}. Researchers noted that deployment mirrors data collection challenges, particularly around representative coverage, equitable device access, and affordability—all essential to prevent exclusion of certain groups from system benefits (P10, P12; \rec{15}).


\subsection{Technical and Practical Barriers to Fair Practice} While the strategies above illustrate ways researchers attempt to anticipate and mitigate fairness risks in passive sensing, they also highlighted deeper technical and structural barriers that constrain current fair practice in reality. 

\subsubsection{Coarse Labels and Deterministic Models Fail to Capture Human Complexity.} 
Researchers described how fairness is constrained by a mismatch between the richness of human states and the limited representational capacity of current models. While passive sensing can infer increasingly detailed physical behaviors (\eg posture, activity, movement), extending these inferences to emotions or mental health was described as ``\textit{hard to be fair},'' given how deeply these states are shaped by \textit{upbringing}'' and \textit{environment}'' (P2). A recurring concern was that modeling pipelines rely on deterministic classifications and simplified ground truth definitions that compress nuanced experiences into \textit{very limited classes}'' based on research assumptions (P6). Researchers noted that mapping behaviors or symptoms into binary or scaled labels---such as whether someone is \textit{working well or not}''---often \textit{doesn't reflect realistically}'' actual performance or lived experience (P12), creating misleading impressions when deployed. These representational limits were also seen to intersect with study design choices, such as excluding comorbid conditions or focusing on narrowly defined populations. As P7 noted, findings derived from ``\textit{just major depressive disorder}'' are difficult to apply in practice, since real patients ``\textit{rarely}'' present without comorbidities. Collectively, researchers framed these issues as fundamental technical constraints that limit how fairly passive sensing can be designed, validated, and applied in real-world contexts. 

\subsubsection{Richer Data vs. Participation Burden.}  The tension between collecting richer data and minimizing participation burden further exacerbates fairness risks. Researchers emphasized the need for data that is both representative and ecologically valid, particularly when passive sensing systems are deployed in real-world, high-stakes contexts. In practice, however, obtaining such data is difficult. P9 noted that many studies rely on small, demographically narrow samples, while scaling to large, diverse cohorts is often infeasible due to the cost of sensing devices and the burden of sustained data collection, which they described as ``\textit{very expensive}'' and placing an ``\textit{undue burden}'' on participants and patients.  Longitudinal demands compound the problem: because behavior is dynamic, longer observation is needed, but those with fewer resources, cognitive impairments, or high clinical burden are least able to provide it. Reflecting on ICU work, P9 noted ``\textit{it is not possible to collect long-term data},'' and a separate three-month nursing-care study required ``\textit{very precautions with various special measures},'' underscoring how sustaining valid data selectively excludes exactly the populations for whom support is most critical.

\subsubsection{Structural and Institutional Pressure.} 
Beyond technical and practical challenges, researchers emphasized that broader systemic forces shape the possibilities for fair practice. Many noted that the field often prioritizes accuracy over fairness---not due to lack of awareness, but because community norms and incentive structures reward performance metrics, speed, and novelty. P4 admitted that their team emphasized model accuracy over fairness ``\textit{to get attention or to get the paper accepted},'' despite knowing their models did not generalize to dementia patients. Similarly, P2 reflected that ``\textit{as an ML researcher, you rarely think about these things... you're going after performance}.'' Others echoed that fairness was often relegated to ``\textit{limitations and future works}'' (P5), or treated as secondary to addressing complex health problems such as depression or stress (P11). These dynamics were further reinforced by institutional contexts. P3 described the ``\textit{peer pressure and fast publishing culture}'' in computer science, where ``\textit{everybody's publishing---with LLMs, things are even crazier},'' leaving little time for reflection on fairness. P7 similarly noted that researchers may only take fairness seriously when it is explicitly mandated by publication venues or funders. For others, fairness simply did not align with their research priorities, or was sidelined by organizational deliverables that emphasized engineering robustness over normative reflection (P2, P3, P6). 

\section{Discussion and Conclusion}

In this paper, we contribute to addressing the gap in domain-specific understandings of fairness sources and risks in wellbeing sensing. Below, we examine how our findings both echo and extend existing literature, and we highlight the structural barriers that hinder the translation of fairness insights into current practice, calling for broader ecosystem-level support for fair passive sensing.

\subsection{From Identity-based Metrics to Lifecycle-Situated Fairness}
Much prior fairness work has centered on algorithmic audits that compare model performance across identity groups~\cite{groves2024auditing,pahl2022female,yfantidou2023beyond}. While identity-based audits remain important, our interviews highlight their limits in passive sensing contexts. Researchers consistently described cases where identity-based attributes failed to account for disparities in representation and access, participation burden and error, and procedural understanding and recourse, often producing mis-specified comparison groups and misleading conclusions. Instead, inequities frequently aligned with situational factors---device access, behavioral regularity, comfort with monitoring---that cut across traditional identity categories. This reframing has direct governance implications: existing regulatory frameworks such as HIPAA-governed health data~\cite{act1996health} and the Common Rule~\cite{commonrule} tend to operationalize fairness through demographic categories, meaning that regulatory compliance may be necessary but insufficient for passive sensing systems. A system satisfying demographic parity requirements may still systematically disadvantage participants based on device ownership or digital literacy---factors that current audit mandates do not require operators to surface.

The fairness risks identified in our interviews are tightly coupled to domain-specific characteristics of passive sensing, including indirect behavioral inference, longitudinal data collection, heterogeneous sensing infrastructures, and asymmetric understanding between system builders and participants. These characteristics surface risks such as unverified inference, uneven participation burden over time, compliance-driven data gaps, and misaligned interpretations of behavioral signals---risks that are less visible in more bounded \gls{ml} settings. At the same time, many of these issues reflect broader patterns across \gls{ml} systems, including convenience-driven sampling, context-insensitive preprocessing, and evaluation practices that obscure who bears error and burden. Accordingly, the mitigation strategies articulated by researchers echo established recommendations in ML and HCI---such as transparency, participatory and context-aware design, careful population scoping, and reflexive evaluation---but require adaptation to the realities of passive sensing. Our primary contribution is not to propose new metrics, but to empirically ground these principles in everyday sensing practice by linking fairness risks to situated sources of inequity and tracing how they emerge across the sensing lifecycle. This framing positions passive sensing as a revealing case for lifecycle-situated fairness that both draws from and informs broader fairness work in \gls{ml}.

\subsection{From Awareness to Fair Practice: Structural Barriers and Ecosystem Support}
A striking theme across interviews was a persistent gap between fairness awareness and fairness action. Researchers widely recognized potential harms and expressed strong normative commitments to addressing fairness risks, along with a clear desire for concrete, actionable guidance (\eg checklists, worked examples, or templates) that could be integrated into ongoing projects. However, they consistently described structural and institutional conditions that constrained their ability to act on these concerns in practice. These included venue incentives that prioritize benchmark gains and novelty over participant protections and fairness evaluation; limited time and funding to recruit beyond convenience samples; missing or unavailable subgroup attributes in public datasets; compute and expertise constraints for stratified analyses; and inconsistent review expectations regarding which fairness metrics and artifacts are required.

As with many \gls{ml} systems, passive sensing systems are inherently socio-technical and are often deployed in high-stakes domains such as health and education~\cite{adler2020predicting,zhang2025towards,wang2014studentlife}. Our findings suggest that enabling fair practice in this space cannot be left to individual researchers alone. Instead, fairness requires ecosystem-level support that makes lifecycle-aware action feasible within real research constraints. This includes embedding a minimum viable fairness audit across the research process; establishing clearer expectations from venues and reviewers around fairness analyses and reporting; funding structures that support recruitment beyond convenience samples and longitudinal validation; shared datasets and tooling that document device constraints, missingness, and subgroup limitations; and institutional norms that reward transparency about trade-offs rather than penalizing incomplete or negative results.

Translating these needs into concrete governance mechanisms points to several actionable directions across the ecosystem:

\begin{itemize}
    \item \textbf{Funding bodies} (e.g., NIH, NSF, Wellcome Trust) could require that grants including passive sensing in health or workplace contexts include a pre-registered fairness protocol covering recruitment diversity, consent accessibility, and behavioral inference validation---analogous to existing requirements for sex and gender inclusion in clinical research.
    \item \textbf{Publication venues} and program committees could establish minimum reporting standards for fairness documentation, such as requiring authors to disclose the subgroup composition of training data, the basis for comparison group definitions, and any structural barriers that constrained fairness evaluation, rather than treating these as optional limitations.
    \item \textbf{IRB and ethics review bodies} are well-positioned to extend their mandate beyond privacy and consent to include distributional fairness risks, particularly for longitudinal deployments where monitoring burden accumulates unevenly. This could include requiring pilot-stage dropout analyses and burden assessments as a condition of full-study approval.
    \item \textbf{Deployment institutions} such as hospitals, universities, and employers bear a responsibility that existing accountability frameworks often overlook. When passive sensing is embedded in institutional infrastructure, the deploying organization acquires ongoing obligations to monitor post-deployment performance disparities, provide meaningful recourse mechanisms, and disclose system limitations to affected individuals.
\end{itemize}

Addressing these barriers requires treating fairness not only as a research norm but as a governance obligation---one that demands coordinated action from funders, publication venues, institutional review boards, and the organizations that deploy passive sensing systems in high-stakes settings~\cite{raji2020closing,madaio2020co}. 

\subsection{Limitations and Future Work}

This study has several limitations that point to important directions for future work. First, our findings are based on interviews with 14 researchers and practitioners working primarily in health and wellbeing---consistent with norms for semi-structured interview studies using \gls{ta}~\cite{braun_using_2006}, and interview depth ranged from 70 to 110 minutes. However, the sample is not intended to be statistically representative. Crucially, our analysis relies solely on researcher accounts rather than the lived experiences of data providers, end users, or domain experts. Future research should directly incorporate these stakeholder perspectives and examine how fairness risks and mitigation strategies differ across application areas and governance regimes.

Second, we do not claim that the fairness risks or mitigation strategies identified in this paper are exhaustive or universally applicable. Passive sensing systems are highly heterogeneous, and fairness risks evolve as sensing modalities, inference techniques, and deployment settings change. Future work should explore whether additional fairness risks exist and evaluate how mitigation strategies shape outcomes over time. 

Finally, while we synthesize recurring fairness risks and mitigation strategies across the passive sensing lifecycle, we do not propose a formal framework or checklist. Instead, our contribution is to make visible where fairness breaks down in practice and how researchers navigate these risks under real constraints. Future work could translate these insights into concrete artifacts---such as lightweight fairness checkpoints and lifecycle-aware audit tools---and study how institutional incentives, review practices, and funding structures enable or constrain their adoption.

\section*{Acknowledgments}
We thank our interview participants for generously sharing their time, experiences, and candid reflections on their own research practice. We are also grateful to the anonymous reviewers and senior program committee members, whose feedback substantially improved the framing of this paper.

\bibliography{main}

\clearpage
\subsection{Screening Survey Questions}\label{appdx:screening_survey}
\begin{itemize}
    \item What is your name? 
    \item Are you above 18?
    \item What is your gender?
    \item What is your primary role in passive sensing research?
    \item How many years have you worked in passive sensing research?
    \item What domains do you primarily apply behavioral sensing to? (Select all that apply)
    \begin{itemize}
        \item Health \& Wellbeing Monitoring
        \item Workplace Productivity
        \item Smart Homes \& IoT
        \item Social Behavior \& Human Interaction
        \item Other
        \end{itemize}
    \item Have you ever explicitly considered fairness-related concerns in your passive sensing research?
    \begin{itemize}
        \item If yes, what factors do you consider as potential fairness risks?
    \end{itemize}
    \item What challenges prevent you from explicitly considering fairness in your research?
    \item Do you think fairness concerns extend beyond model training and testing in passive sensing research?
    \item Have you deployed a passive sensing system in the real world?
\end{itemize}

\subsection{Interview Questions}\label{appdx:interview_study}
\begin{itemize}
    \item Part 1: Fairness Definitions \& Awareness
    \begin{itemize}
        \item Could you briefly describe a passive sensing study you conducted? What were the study's goals and methodology?
        \item Could you walk us through the end-to-end pipeline for this study? What considerations did you take into account at each step?
        \item How do you define fairness in the context of passive sensing research? How, if at all, does it differ from fairness in other domains? 
        \item Have you used sensing data to infer human behaviors? If so, what methods or approaches did you use to infer those behaviors?
        \item Have there been times when inferred behaviors did not reflect the actual situation?
        \begin{itemize}
            \item If yes, how did you recognize the mismatch?
            \item If not, did you verify and find no discrepancies, or has this not been examined?
        \end{itemize}
        \item Have you considered or researched fairness in your previous studies?
        \begin{itemize}
            \item If not, what challenges or barriers prevented you from doing so?
        \end{itemize}
        \item In your view, what should the field do to improve awareness of fairness in passive sensing research?
    \end{itemize}

    \item Part 2: Fairness Risks \& Mitigation Strategies
    \begin{itemize}
        \item Which step(s) in the pipeline present potential fairness risks? What specific concerns are associated with each step?
        \item What methods or metrics do you find useful for evaluating fairness in passive sensing research?
        \item Can you describe a specific instance where fairness issues emerged in your research (or in a study you examined)? How did you address them, if at all?
        \item What strategies should be used to mitigate biases arising from the longitudinal and dynamic nature of passive sensing?
        \item Have you deployed your passive sensing systems in the real world?
        \begin{itemize}
            \item If so, did you encounter any fairness concerns after deployment? What were they, and how did you respond?
        \end{itemize}
    \end{itemize}

    \item Part 3: Opportunities \& Feedback
    \begin{itemize}
        \item Considering a complete passive sensing study lifecycle, what fairness issues and other potential harms should researchers consider and assess at each step?
        \item Below is a proposed framework. Do you have any feedback or suggestions for improvement?
    \end{itemize}
\end{itemize}

\subsection{Details about Participants}
\begin{table*}[!t]
\sffamily
\rowcolors{2}{gray!10}{white}
\caption[Summary of interviewed participants.]{Summary of interviewed participants. The table below provides an overview of the 14 participants included in our study. For each participant, we report self-identified gender, current professional role, country of their workplace, years of experience in behavioral sensing, and primary domain(s) of expertise. Continent of workplace is shown to preserve anonymity.}\label{tab:participant_summary}
\renewcommand{\arraystretch}{1.2}
\footnotesize
 \resizebox{\textwidth}{!}{
\begin{tabular}{C{0.06\textwidth} L{0.09\textwidth} L{0.23\textwidth} L{0.16\textwidth} L{0.13\textwidth} N{0.45\textwidth}}
\hline
\multicolumn{1}{r}{{\textbf{\makecell[c]{PID}}}} & \multicolumn{1}{c}{{\textbf{\makecell[c]{Gender}}}} & \multicolumn{1}{c}{{\textbf{\makecell[l]{Current Role}}}} & \multicolumn{1}{c}{{\textbf{\makecell[c]{Continent of \\ Workplace}}}} & \multicolumn{1}{c}{{\textbf{\makecell[c]{Years of \\ Experience}}}} & \multicolumn{1}{c}{{\textbf{\makecell[c]{Primary Domain(s) of Expertise}}}} \\\hline
PID1 & Female & Industry Researcher & Oceania/Asia& 7+ years & Health $\&$ Well-being Monitoring, Privacy, Smart Homes $\&$ IoT\\
PID2 & Male & Faculty/PI & Oceania/Asia& 4-6 years & Health $\&$ Well-being Monitoring, Social Behavior $\&$ Human Interaction\\
PID3 & Male & Industry Researcher & North America & 7+ years & Health $\&$ Well-being Monitoring \\
PID4 & Female & PhD Student & Oceania/Asia& 4-6 years & Health $\&$ Well-being Monitoring, Smart Homes $\&$ IoT, Social Behavior $\&$ Human Interaction \\
PID5 & Male & Postdoctoral Researcher & Oceania/Asia & 4-6 years & Health $\&$ Well-being Monitoring, Causal Inference\\
PID6 & Female & PhD Student & Oceania/Asia & 1-3 years & Health $\&$ Well-being Monitoring, Social Behavior $\&$ Human Interaction\\
PID7 & Male & PhD Student & North America & 4-6 years & Health $\&$ Well-being Monitoring, Social Behavior $\&$ Human Interaction\\
PID8 & Male & Industry Researcher & North America & 7+ years & Health $\&$ Well-being Monitoring, Workplace Productivity, Smart Homes $\&$ IoT, Social Behavior $\&$ Human Interaction \\
PID9 & Female & Faculty/PI & Oceania/Asia & 7+ years & Health $\&$ Well-being Monitoring, Workplace Productivity, Smart Homes $\&$ IoT, Social Behavior $\&$ Human Interaction\\
PID10 & Male & PhD Student & North America & 4-6 years & Health $\&$ Well-being Monitoring, Smart Homes $\&$ IoT \\
PID11 & Male & PhD Student & North America & 4-6 years & Health $\&$ Well-being Monitoring, Social Behavior $\&$ Human Interaction\\
PID12 & Male & PhD Student & Oceania/Asia& 1-3 years & Workplace Productivity, Social Behavior $\&$ Human Interaction\\
PID13 & Male & Postdoctoral Researcher & North America & 7+ years & Health $\&$ Well-being Monitoring, Workplace Productivity, Social Behavior $\&$ Human Interaction\\
PID14 & Female & PhD Student & Oceania/Asia & 1-3 years & Health $\&$ Well-being Monitoring \\ \hline
\end{tabular}}
\end{table*}


\end{document}